\documentstyle[aps,prl,draft]{revtex}
\begin{document}

\draft
\title{Novel type of orbital ordering: complex orbitals in doped Mott
insulators}
\author{D.~Khomskii}
\address{Laboratory of Solid State Physics, Groningen University,\\
Nijenborgh 4, 9747 AG Groningen, The Netherlands}
\maketitle
\widetext

\begin{abstract}
An orbital ordering, often observed in Mott insulators
with orbital degeneracy, is usually  supposed to disappear with doping,
e.g.\ in the ferromagnetic metallic phase of manganites.  We propose
that
the orbital
ordering of a novel type may exist in such situation: there may occur
ferro.\ orbital
ordering of {\it complex orbitals} (linear superposition of basic
orbitals
$d_{x^2-y^2}$ and $d_{z^2}$ with complex
coefficients).  Despite the perfect orbital ordering, such state
still retains cubic symmetry and thus would not induce any structural
distortion.  This novel state can resolve many
problems in the physics of CMR manganites and can also exist
in other doped Mott insulators with Jahn-Teller ions.
\end{abstract}
\pacs{PACS numbers: 75.30V, 71.70E}

\baselineskip 24pt

Mott insulators often display various types of ordering
connected with localized electrons.  In particular, in the presence of
an orbital degeneracy, very often an orbital ordering
occurs besides the magnetic (spin) one.
Classical examples are the systems with doubly-degenerate
$e_g$-electrons, e.g.\ KCuF$_3$ or MnF$_3$~\cite{Kugel&Khomskii}.
At present special attention is paid to the manganites
La$_{1-x}M_x$MnO$_3$ ($M=\rm Ca,Sr$) with the colossal
magnetoresistance (CMR).  The undoped material LaMnO$_3$ is a Mott
insulator with the orbital ordering.
What becomes of this ordering in doped materials and which role the
orbital degrees of freedom do play in the CMR phase, is a matter of
active
investigation and of hot debate.
On the one hand it is usually
assumed that the orbital ordering (at least the long-range one)
disappears in a ferromagnetic metallic (FM) phase ($0.16<x<0.5$)
displaying CMR\null.  This is in particular concluded from
the disappearance of lattice distortion typical for the orbitally
ordered states.  On the other hand the attempts to
explain
transport properties in this regime using only electron--spin
scattering in the framework of the standard double-exchange model
failed~\cite{Millis}, which forced Millis and coworker to suggest
that the orbital degrees of freedom are still important in this
regime, e.g.\ producing Jahn-Teller polarons. Nevertheless the
orbitals were still assumed to be disordered in the FM phase
below~$T_c$.

Until now in all the numerous studies of the cooperative Jahn-Teller
effect and orbital ordering only one class of solutions was
considered: occupied orbitals were always assumed to be certain
linear combinations of the basic orbitals
$d_{z^2}\sim\smash{\frac{1}{\sqrt6}}(2z^2-x^2-y^2)$ and
$d_{x^2-y^2}\sim\smash{\frac{1}{\sqrt2}}(x^2-y^2)$ of the form
\begin{equation}
|\theta\rangle=\cos\frac\theta2\,|z^2\rangle+\sin\frac\theta2\,|x^2-
y^2\rangle\;,
\label{eq1}
\end{equation}
i.e.\ linear superposition with real coefficients.  Such orbitals
give an asymmetric (qua\-d\-ru\-po\-lar) distortion of the electron
density at
a given site.  Consequently, any ordering
involving orbitals of the type~(\ref{eq1}), be it of ``ferro'' or
``antiferro'' type, corresponds to a decrease of the symmetry of
charge distribution and is accompanied (or driven) by the
corresponding lattice distortion.

There exists however yet another possibility: there may exist {\it
orbital ordering without any lattice distortion}.  From the basis set
of doubly-degenerate orbitals we can also form a linear superposition of

the type~(\ref{eq1}) but {\it with complex coefficients}.  The simplest
example is the normalized functions
\begin{equation}
|\pm\rangle=\frac{1}{\sqrt2}\bigl(|z^2\rangle\pm
i|x^2-y^2\rangle\bigr)\;.
\label{eq2}
\end{equation}
We can form an orbital ordering with the occupied orbitals of such
type, e.g.\ the ferro.\ orbital ordering (FOO) with the orbitals
$|{+}\rangle$ at each site.
It can be easily
checked that the electron charge distribution in this state is
perfectly symmetric, it is the same along $x$, $y$ and $z$-directions,
so that the net symmetry of a crystal would remain cubic.
Correspondingly, the effective electron hopping matrix elements are
isotropic, $t^x=t^y=t^z=t^*$, as
well as superexchange interaction $J^x=J^y=J^z=2t^{*2}/U$.
(Here we have in mind the traditional description of the strongly
interacting electrons by the Hubbard-type model which in case of orbital

degeneracy is actually a degenerate Hubbard model \cite{Kugel&Khomskii}
with the on-site repulsion $U$ and electron hopping between different
orbitals, to be specified below.)
We suggest that just such an ordering is realized
in the FM phase of CMR manganites.  If true, this would imply that
both the spin and orbital degeneracies are lifted in the ground state
in these systems which consequently are {\it ferromagnetic in both the
spin and orbital variables} (but with ``strange''
orbitals~(\ref{eq2})).

Why have such states not been considered before?  Apparently it is
connected with two factors.  On one hand, in most real materials with
Jahn-Teller ions studied up to now there always occurred orbital
ordering
accompanied by lattice distortion.  On the other hand, in the
theoretical description of these phenomena only the operators
corresponding to an ordering of conventional type appear in the
effective
Hamiltonian (see below), so that it seemed that other possibilities are
never realized in practice \cite{Korovin&Kudinov,footnote}.  Most
probably, indeed only the ``real'' solutions~(\ref{eq1}) can be realized

in
stoichiometric compounds with Jahn-Teller ions.  The situation however
may
be different in doped materials --- and that is what we propose here.

It is well known that the electron (or hole) motion in systems with
strongly correlated electrons is hindered by an antiferromagnetic
background and is much facilitated if the background ordering becomes
ferromagnetic (see e.g.~\cite{Khomskii&Sawatzky}): in nondegenerate
Hubbard model it gives rise to Nagaoka's ferromagnetism~\cite{Nagaoka},
and
in systems with two types of electrons --- to
ferromagnetism due to double exchange.  The energy gain in such a
ferro.\
state as compared to the antiferro.\ or paramagnetic (disordered) one
is of the order of the electron bandwidth, or of the hopping matrix
element
$t$ per doped charge carrier: $\Delta E=-ctx$ ($c$ is some constant of
order~1),
whereas the energy loss is of the order of $J$~$(\sim t^2/U)\cdot(1-x)$.

Thus, as argued already in \cite{Nagaoka}, the saturated ferromagnetic
state is realized for $x>x_c\sim J/t\sim t/U$.  For smaller $x$ the
inhomogeneous phase separated state with ferromagnetic droplets
in an antiferomagnetic matrix can be realized~\cite{Visscher}.

In exact analogy to this case we should expect that due to the same
mechanism a ferro.\ orbital ordering will be established in doped
strongly
interacting Mott insulators with orbital degeneracy.
Which particular orbitals would be stabilized at
that, may depend on the particular situation; we suggest that it may
be the complex orbitals of the type~(\ref{eq2}).

A convenient way to
describe the orbital ordering of double-degenerate $e_g$-orbitals is
to introduce pseudospin variables $\tau_i$ such that e.g.\ the orbital
$d_{z^2}$ corresponds to $\tau^z=+\frac12$, and the
orthogonal orbital $d_{x^2-y^2}$ --- to $\tau^z=-\frac12$.
The orbital states~(\ref{eq1}) considered until now are parametrized
by the angle $\theta$ in the
$(\tau^z,\tau^x)$-plane.
Cubic symmetry is reflected in the $\theta$-plane in
$\frac{2\pi}3$ symmetry: the state $\theta=\frac{2\pi}3$ corresponds
to the orbital $|x^2\rangle\sim(2x^2-y^2-z^2)$ and
$\theta=-\frac{2\pi}3$ --- to $|y^2\rangle\sim(2y^2-x^2-z^2)$; these
are more or less the orbitals occupied in two sublattices in the
undoped LaMnO$_3$.  When we dope the system with orbital
ordering, the motion of the charge carriers
would initially (for small $x$) occur on a background of this orbital
ordering.  Similar to the hole motion on an
antiferromagnetic spin background, the ``antiferro'' orbital ordering
would hinder the motion of a hole and would reduce its
bandwidth.
One can indeed check that, by making the orbital order ferromagnetic,
e.g.\ occupying the same orbital at each site, for instance
$|z^2\rangle$ or $|z^2-\nobreak y^2\rangle$, we would increase the
bandwidth
and correspondingly decrease kinetic energy of holes; the mathematical
details are presented below.  However such a FOO seems to contradict
experimental observations.  Thus e.g.\
$|z^2\rangle$-fer\-ro\-mag\-ne\-tism
would lead to a tetragonal distortion of the lattice with $c/a>1$ and
to a strong anisotropy of transport properties;
but nothing like that is observed experimentally.
Probably that is why this possibility, which by analogy with the spin
case seems quite natural, is never considered for CMR
phase of manganites, and it is usually assumed that the orbital ordering

is simply lost in the FM phase leading e.g.\ to an orbital
liquid~\cite{Nagaosa}
However if at such a FOO the {\it complex} orbitals (\ref{eq2})
would be occupied --- there would be no contradiction with the
structural data, and still the motion of doped charge carriers would
be unhindered.  This is the main idea of the present paper.
In the pseudospin language introduced above the state~(\ref{eq2}) is
an eigenstate of the operator $\tau^y$ which also exists in the
algebra of $\tau=\frac12$ operators.

For strongly interacting electrons with one electron per site one can
describe the effective
spin and orbital interaction by the Hamiltonian which schematically has
the
form~\cite{Kugel&Khomskii}
\begin{equation}
{\cal H}=\sum J_s(\vec s_i\vec
s_j)+J_\tau(\tau_i\tau_j)+J_{s\tau}(\vec s_i\vec
s_j)(\tau_i\tau_j)\,,
\label{eq3}
\end{equation}
where the first and the third terms are due to superexchange,
$J_{exch}\sim J_s\sim J_{s\tau}\sim t^2/U$,
and in the second term also the Jahn-Teller interaction of degenerate
orbitals with the lattice contribute, $J_\tau=J_{exch}+J_{\rm JT}$
The interaction (\ref{eq3}) is in general anisotropic with respect to
$\tau$-operators and usually it does not contain terms with~$\tau^y$.
Consequently $\tau^y$-states do not appear in the mean-field
approximation
 which is nearly always used to
treat insulators with orbital degeneracy.  But it does not mean that
such states cannot appear in certain situations, e.g.\ for doped
systems.

Using
the specific form of the $e_g$-orbitals and of the corresponding
hopping integrals between different orbitals in different directions,
one can easily calculate the one-electron band structure for
different types of orbital ordering.  Denoting the hopping between
$z^2$-orbitals for a pair along $z$-direction as~$t$, we have:
$t^z_{z^2,z^2}=t$,  $t^{x,y}_{z^2,z^2}=t/4$,
$t^{x,y}_{x^2-y^2,x^2-y^2}=3t/4$,
$t^{x,y}_{z^2,x^2-y^2}=\pm\sqrt3\,t/4$,
$t^x_{z^2,x^2}=t^{x,y}_{x^2,y^2}=t/2$,
etc., see e.g.~\cite{Kugel&Khomskii}

Generally speaking, in contrast to spin case (see
\cite{Khomskii&Sawatzky} for discussion and references), an
antiferro.\ orbital ordering, e.g. that in LaMnO$_3$, does not
completely suppress the hole motion even if one ignores quantum
effects: the hole can always hop to a neighbouring site without
destroying
orbital order along its trajectory.  This is connected with the fact
that the nondiagonal hopping in orbital channel is in general allowed,
and
pseudospin projection $\tau^z$ is not conserved during hopping:
the nondiagonal hopping integrals $t_{z^2,x^2-y^2}$ are in general
nonzero.  Thus the hole can here move without leaving
a ``trace'' of wrong spins~\cite{Bulaevskii} and, consequently,
there will be no ``confinement''.  However the bandwidth of this
coherent motion without disturbing the background orbital ordering
is reduced as compared to the bandwidth for FOO\null.  Thus,
using the values of hopping integrals $t_{\alpha,\beta}^{\langle
ij\rangle}$
given above, one obtains for the undoped LaMnO$_3$ (alternation of
$x^2$ and $y^2$ orbitals) the
spectrum
\begin{equation}
\textstyle
\varepsilon(k)=-2\left[\frac12t(\cos k_x+\cos
k_y)+\frac14t\cos k_z\right]
\label{eq4*}
\end{equation}
so that the minimum energy of the hole (the bottom of the band) will
be $\varepsilon_{\it min}=\varepsilon(k{=}0)=-2.5t$.
If however we would make a FOO, e.g.\ occupying at
each site $z^2$-orbital, the spectrum will be
\begin{equation}
\textstyle
\varepsilon^{z^2\hbox{\scriptsize\it-ferro}}(k)=-2t\left[\frac14(\cos
k_x+\cos
k_y)+\cos k_z\right]
\label{eq5*}
\end{equation}
and $\varepsilon^{z^2\hbox{\scriptsize\it-ferro}}_{\it min}=-3t$.
Thus we indeed see that the energy gain in the FOO state is of order
$\sim t$.
The same minimum energy is reached also for $(x^2-y^2)$-ferro.\
ordering and for FOO with any orbital of the type~(\ref{eq1}).
From~(\ref{eq1}) one can easily obtain that
\begin{equation}
\textstyle
t_{\theta,\theta}^z=\langle\theta|\hat
t\,^z|\theta\rangle=t\cos^2\frac\theta2;\qquad
t_{\theta,\theta}^{x\,{/}\,y}=\frac
t4(\cos\frac\theta2\pm\sqrt3\,\sin\frac\theta2)^2\;,
\label{eq9n}
\end{equation}
And one obtains that the bottom of the spectrum
with the hopping integrals given by~(\ref{eq9n}) does not depend on
$\theta$ and coincides with the value~$-3t$.

But exactly the same $\varepsilon_{\it min}$ is also reached for the
$\tau^y$-ferromagnetism, where the state~(\ref{eq2}) is occupied at
each site.  Using the values of $t$'s presented above, one obtains
that $t_{\tau^y,\tau^y}=\frac12t$ in all three directions, and
\begin{equation}
\varepsilon^{\tau^y\hbox{\scriptsize\it-ferro}}(k)=-t(\cos k_x+\cos
k_y+\cos k_z)
\end{equation}
so that the spectrum is indeed isotropic, and the minimum energy is
also equal to $-3t$.

Note that, similarly to the nondegenerate Hubbard model, we can use the
simple dispersion relations (\ref{eq4*}), (\ref{eq5*}) only for the
electron motion which does not destroy the background ordering.  The
energy
gain in the FOO state is again, similarly to the Nagaoka
case~\cite{Nagaoka},
$\sim tx$, and the energy loss is $\sim J$ (in orbital sector $\sim
J_\tau+J_{s\tau}$).
Here again we obtain the totally ferromagnetic state both in spins and
in orbitals
for $x>x_c\sim J/t$, and one can have phase separation for
$x<x_c$.

As we saw above, for small $x$ different FOO states are equivalent as
to the gain in kinetic energy.  On the other hand we can see that
$\tau^y$-ferro.\ state is favourable as to the loss of the effective
exchange energy~(\ref{eq3}).  Indeed, the effective Hamiltonian
(\ref{eq3})
contains the orbital operators in combinations
$\tau^z_i \tau^z_j$ , $\tau^x_i \tau^x_j$ and $\tau^z_i \tau^x_j$
with positive (antiferromagnetic) coeficients
$J_{\tau}$, $J_{s\tau}$~\cite{Kugel&Khomskii}.
In the ground state of the undoped system a certain spin and orbital
order is
realised which minimizes the total energy --- and it will be an
antiferro.\
ordering of real orbitals~(\ref{eq1}).
 If however we {\it force} our system to be
ferromagnetic both in spins and in orbitals, e.g., as argued above, by
doping,
we strongly increase this exchange energy --- and we can minimize this
energy loss by chosing FOO of $\tau^y$-type.

One can check directly that the superexchange part of the energy is
indeed smaller in the $\tau^y$-FOO state as compared e.g.\ with the FOO
state of real orbitals.  Thus,
taking the standard expression for the exchange integrals $J=2t^2/U$
and calculating exchange constants with the proper values of $t$,
 we obtain e.g.\ that for the
$z^2$-FOO the energy of the ferromagnetic state (per site)
in mean-field approximation can be written  in the form:
\begin{equation}
E_{\it
ferro}^{z^2\hbox{\scriptsize-FOO}}=\frac94\frac{t^2}{U}
+ C\;,
\label{eq11}
\end{equation}
where $C = - 3 \frac{t^2}{U}$.
The same value of $E_{\it ferro}$ we would obtain for $(x^2-y^2)$-FOO
and for all other FOO states with arbitrary state $|\theta\rangle$ of
the type~(\ref{eq1}): again one can easily show that with the values
of hopping integrals $t_{\theta\theta}^\alpha$~(\ref{eq9n}) the
magnetic energy of both ferro- and antiferromagnetic states does not
depend on $\theta$ and is given by (\ref{eq11}).
Similar treatment of the $\tau^y$-FOO state shows however
that the net exchange energy in this state is lower:
\begin{equation}
E_{\it
ferro}^{\tau^y\hbox{\scriptsize-FOO}}=\frac32\frac{t^2}{U}\,
+ C\;,
\end{equation}
where $C$ is the same as in Eq.(\ref{eq11}).
Similarly, the intersite Jahn-Teller energy in FOO state with real
orbitals
is positive, $\sim J_{\rm JT}$, whereas it is zero for the $\tau^y$-FOO
state; again from this point of view the $\tau^y$-FOO state wins against

all other possible FOO states with real orbitals.

As the electronic energy of doped charge carriers coincides for the
states~(\ref{eq1}) and~(\ref{eq2}),
and the magnetic energy of the latter is lower, one
can conclude that the proposed ferromagnetic state with $\tau^y$-FOO
(occupation of complex orbitals~(\ref{eq2})) will be the best among
all possible ferro.\ ordered states --- and these, we argue, may be
stabilized by the usual Nagaoka mechanism.  This state
can possibly be realized in the ferromagnetic metallic
phase of CMR manganites.  The finite band filling would increase
electron kinetic energy somewhat faster for $\tau^y$-ordering as
compared e.g.\ to $z^2$-FOO or $(x^2-y^2)$-FOO (the energy
spectrum in the latter cases is anisotropic and the density of
states at the band edge is higher,
cf.~\cite{vandenBrink}).  Consequently the
$\tau^y$-FOO may be destabilized at still higher doping levels.

The state
with $\tau^y$-FOO need not be necessarily homogeneous: as in the
pure spin case, for small doping there may exist a tendency towards
phase
separation  due to an
instability of the homogeneous canted spin
state~\cite{Visscher,Yunoki}.  But, as
follows from the arguments given above,
the structure of the hole-rich regions can well be of $\tau^y$-FOO
type.

>From the point of view of symmetry the proposed $\tau^y$-FOO
corresponds
to the $A_{2g}$ (or $\Gamma_2$)-representation of the cubic group
\cite{Abragam}.
The corresponding order parameter is
\begin{equation}
T_{xyz}=\langle{\cal S}l_xl_yl_z\rangle\,,
\label{new2}
\end{equation}
where $\langle\;\rangle$ is the ground state average, ${\cal S}$ denotes

symmetrization and $l_x$, $l_y$, $l_z$ are the corresponding components
of the momentum operator $\hat l=2$.  This follows from the group
symmetry,
and we checked by direct calculation that it is the lowest nonzero
average in our state.  One can show that in the basis $|z^2\rangle$,
$|x^2-y^2\rangle$
$T_{xyz}$ is indeed proportional to the $\tau^y$-Pauli matrix.  From
eq.~(\ref{new2}) we see that the order parameter in our state is a
magnetic octopole, cf.~\cite{Korovin&Kudinov,footnote}.  It is in
principle possible, although not easy, to observe such ordering
experimentally.  Probably the most promising would be the resonance
experiments like NMR or circular X-Ray dichroism.  One can also show
that this state should have a piezomagnetism.

It is not apriori clear whether this $\tau^y$-FOO and the ferromagnetic
spin ordering would occur at the same temperature.  The direct product
of corresponding irreducible representations $A_2$ and $T_1$ does not
contain unit
representation $A_1$ \cite{Abragam}, so that there will be in general
no linear coupling between these order parameters.  However if the
rhombohedral
distortion $T_{2g}$, present in manganites at high temperatures
(S.~W.~Cheong, private
communication), would still exist at low temperatures, then there will
exist
linear coupling of these three types of ordering ($A_2\times T_1\times
T_2\in A_1$) and
the $\tau^y$-FOO and the ferromagnetic ordering will occur at the
same~$T_c$.
This could help to
resolve the problem pointed out in Ref.~\cite{Millis}, that the
change of resistivity through $T_c$ in purely spin case is too small:
one would get an extra scattering in a ``para'' phase --- an
orbital-disorder
scattering --- in addition to the spin-disorder one.
(Alternatively one can speak of the increase of resistivity above
$T_c$ not due to orbital fluctuations, but, again in analogy with the
spin case~\cite{Bulaevskii,Brinkman&Rice}, due to the
effective narrowing of the band in the orbitally disordered phase.)

Summarizing, we suggest in this paper that, similarly to magnetic
ordering, doping of orbitally degenerate Mott insulators
containing Jahn-Teller ions could stabilize the ferro.\ orbital
ordering of special type, the occupied orbitals being ``complex'' ---
i.e.\ the linear superpositions of basic orbitals $d_{x^2-y^2}$ and
$d_{z^2}$ with complex coefficients.  Such orbitals, despite perfect
ordering, do not induce any structural distortion.  At the same time
the motion of charge carriers on such ordered background is completely
free, and there is no extra band narrowing;
it is just this factor that favours such ferro.\ orbital ordering.
Thus the origin of the ferro.\ orbital ordering in doped systems is
the same as that of the Nagaoka ferromagnetism in a partially-filled
Hubbard model.
The ferro.\ ordering of complex orbitals gives the same band
energy as the ordering of conventional real ones but has lower
exchange and Jahn-Teller
energy; therefore the ferro.\ ordering of complex orbitals may be
preferable.  This  idea may explain the main properties of the
colossal magnetoresistance  manganites in the most interesting
ferromagnetic metallic concentration range: in this picture they are
perfectly ordered with respect to both spin and orbitals, but with
the {\it complex} orbitals occupied at each site.  In particular, this
could help to
explain the sharp drop of resistivity below $T_c$, with the crystal
structure remaining undistorted.  Similar states
may in principle exist also in other doped Mott insulators
with Jahn-Teller ions which constitute a large interesting class of
magnetic materials with very specific properties.

\smallskip
I am grateful to G.~A.~Sawatzky and M.~V.~Mostovoy for useful
discussions.  I am also grateful do D.~Cox who pointed out a possible
importance of the rhombohedral distortion in manganites.
This work was supported by the Netherlands Foundation FOM,
by the European Network~OXSEN and by the project INTAS--97~0963.

\end{document}